\documentclass[conference]{IEEEtran}
\usepackage[utf8]{inputenc}
\IEEEoverridecommandlockouts
\usepackage{cite}
\usepackage{amsmath,amssymb,amsfonts}
\usepackage{algorithmic}
\usepackage{graphicx}
\usepackage{textcomp}
\usepackage{xcolor}
\usepackage[symbol]{footmisc}
\usepackage{float}
\usepackage{multirow}
\usepackage{makecell}
\usepackage{siunitx}
\usepackage{wrapfig}
\usepackage{subcaption}
\usepackage[font=small]{caption}
\def\BibTeX{{\rm B\kern-.05em{\sc i\kern-.025em b}\kern-.08em
    T\kern-.1667em\lower.7ex\hbox{E}\kern-.125emX}}
\begin{document}

\title{Integrated Switched Capacitor Array and Synchronous Charge Extraction with Adaptive Hybrid MPPT for  Piezoelectric Harvesters}

\author{\IEEEauthorblockN{Pramit Karmakar\textsuperscript{*} }
\IEEEauthorblockA{\textit{Department of Electronics and}\\
\textit{Communication Engineering} \\
\textit{Manipal Academy of Higher Education}\\
Manipal, India \\
pramitkarmakar65@gmail.com}
\and
\IEEEauthorblockN{Siddharth B\textsuperscript{*}}
\IEEEauthorblockA{\textit{Department of Mechatronics} \\
\textit{Manipal Academy of Higher Education}\\
Manipal, India \\
siddharthazad07@gmail.com}
\and
\IEEEauthorblockN{\textsuperscript{}Chinmay Murlidhar Kadnur Rao}
\IEEEauthorblockA{\textit{Department of Electronics and}\\
\textit{Communication Engineering}\\
\textit{Manipal Academy of Higher Education}\\
Manipal, India \\
chinmaykrao2@gmail.com}\thanks{\textsuperscript{*}These authors contributed equally to this work.}
}

\maketitle

\begin{abstract}

Energy Harvesting technologies will play a fundamental role in the development of the next generation of electronic systems as well as in advancing the development of sustainable infrastructure. One of the critical challenges in EH is utilizing ambient vibrations to harvest energy. Piezo Energy Harvesting, which uses ambient vibrations, is a promising technology in energy harvesting and a self-powered technology. However, it suffers from several practical challenges. Some of these challenges include narrow bandwidth, non-linearity, and impedance mismatch, among others. This paper presents a novel, simulated Piezo Energy Harvesting (PEH) framework that addresses some of these challenges. The proposed model is designed to be adaptive and effective against the inherent non-linearity of PEH. This detailed model covers a non-linear piezo, Synchronous Electric Charge Extraction (SECE), Hybrid Maximum Power Point Tracking (MPPT) and a Switched Capacitor Array (SCA). The SECE extracts the maximum charge accumulated on the piezo every time the piezo reaches the mechanical extremum. The Bouc-Wen model has been used to establish nonlinearity in the system. The hybrid MPPT exhibits significant improvement over conventional P\&O, while the SCA-tuned system demonstrates resilience against variable frequency input.
\end{abstract}


\section{Introduction}
Energy Harvesting is the process of capturing ambient energy — such as solar, thermal or mechanical — and converting it into usable electrical power. Among mechanical modalities, Piezoelectric Energy Harvesting (PEH) is particularly attractive for low‑power, self‑powered sensors due to its direct mechanical‑to‑electrical transduction \cite{pradeesh2022review}.
However, PEH systems still have not seen large-scale commercial adoption due to the three interdependent barriers:
\begin{enumerate}
 \item Damping: It is inherent in PEH since every time we extract electricity from it, some energy is lost from the overall structure, and it must see some effect on its dynamics. Damping can therefore significantly decrease the efficiency of PEH \cite{sodano2004review}.
 \item Narrow Bandwidth: Fixed‑tuned harvesters only work effectively at one resonant frequency, yet ambient vibrations can shift by ±10Hz or more.
 \item Nonlinearity: Piezo materials exhibit hysteresis and stiffness variations under large strain, invalidating simple linear spring models. 
\end{enumerate}

In dealing with these issues, we see more advanced and specialised techniques such as Synchronous Electric Charge Extraction (SECE)\cite{lefeuvre2005piezoelectric}  and Synchronized Switch Harvesting on Inductor(SSHI)\cite{wang2021novel}, which appear quite effective in dealing with damping and non-linear response of Piezo. The SECE is regarded as a robust technique as it extracts all the accumulated charge on the Piezo at the extremum of the amplitude of mechanical vibration, and hence it is not load dependent. SSHI on the other hand, can be regarded as a specialised MPPT technique for PEH.

Although each of these techniques has been successfully implemented in further optimizing individual aspects of PEH, negligible work has been done in integrating them in one cohesive architecture that would:

\begin{itemize}
    \item Accurately model and reproduce nonlinear  losses
    \item Extract energy with minimal added damping,
    \item Continuously track and hold the MPP without interrupting energy flow, and
    \item Retune its resonance electrically to follow ambient frequency drift.
\end{itemize}

To fill this gap, the present study develops a fully integrated simulation‑based architecture that meets these requirements:

\begin{enumerate}
    \item Nonlinear Transducer Modeling with Bouc–Wen Hysteresis: Embedding a Bouc–Wen differential model captures amplitude‑dependent restoring forces and per‑cycle energy loss in software, giving the controller the same dynamics as the hardware.
    \item Event‑Driven SECE Extraction: The system waits for each mechanical extremum, then briefly closes an inductor switch to transfer nearly all available charge from an intermediate capacitor into the main storage element, achieving high extraction efficiency and low continuous damping.
    \item Hybrid MPPT (Pseudo‑FOCV Jump + Adaptive P\&O): A coarse repositioning uses the storage capacitor voltage as a pseudo open‑circuit voltage jump, then an adaptive P\&O adjusts its step size based on recent power changes, combining fast disturbance recovery with fine convergence at the true maximum power point.
    \item Digitally Controlled Switched‑Capacitor Array (SCA): A bank of discrete nanofarad‑range capacitors is switched in or out under digital control, alters the piezo’s effective capacitance and thus its resonant frequency. A pre‑characterized lookup table maps sensed frequency to ideal capacitance; the controller then selects the nearest realizable combination, achieving ±10Hz of continuous resonance tracking.
\end{enumerate}

The following sections outline this method's core components, including the SECE, MPPT, and SCA. The results exhibit the robustness of this system in dealing with fluctuating vibrations as well as the efficiency of the proposed Hybrid MPPT.

\section{Literature Review}

The literature review explores some key advancements in the field of PEH such as the Piezoelectric effect, MPPT and Switched Capacitor Arrays(SCA). 

\subsection{Piezoelectric effect and Piezoelectric energy harvesting}
The piezoelectric effect is a phenomenon wherein certain materials generate an electric charge when subjected to mechanical stress \cite{zhang2011piezoelectric}.

In \cite{williams1996analysis}, the authors demonstrate one of the earliest models to harvest energy from ambient vibrations. The model uses an electromagnetic transducer. Since then, there has been a lot of work done to solve practical problems using PEH, such as using the backpack to generate significant amounts of energy without putting any additional pressure on the user \cite{feenstra2008energy}. 

Various modeling approaches have been developed to mitigate the inherently narrow bandwidth limitation of PEH\cite{tang2010toward}. The models can roughly be divided into mechanical methods, magnetic methods, and piezoelectric methods, which are used to change the resonant frequency as needed. This paper proposes an integrated model of PEH. The model tunes the resonance using an electrical method by digitally controlling an SCA to widen the bandwidth of the harvester.

\subsection{Nonlinear Modeling of Piezoelectric Dynamics}
Early PEH models typically approximated the cantilever as a linear spring–mass–damper system. However, real piezoelectric materials exhibit amplitude‑dependent stiffness, saturation, and energy loss due to internal hysteresis when subjected to large strains. To better represent these phenomena, nonlinear models such as the Duffing oscillator (for stiffness variation) and the Bouc–Wen model (for rate and history‑dependent hysteresis) have been developed~\cite{jia2019review,boucwen2009nonlinear}. 

Incorporating such nonlinearities into simulation frameworks has been shown to improve accuracy, especially when estimating power output under realistic vibration profiles. In particular, the Bouc–Wen model captures loop-like force–displacement behavior and per-cycle memory loss—features critical to sizing energy harvesters and designing effective extraction circuits.

\subsection{Charge Extraction Circuits}
Continuous rectification of the piezoelectric voltage imposes significant electrical damping, reducing net energy yield~\cite{lefeuvre2005piezoelectric}. To overcome this, event‑driven extraction circuits have been developed.

Synchronous Electric Charge Extraction (SECE) divides each vibration cycle into two distinct phases: an unloaded “free‑oscillation” period, during which the piezo element charges an intermediate capacitor, followed by a swift LC resonance at the mechanical extremum that transfers nearly all stored charge into the main storage capacitor in a single, high‑efficiency pulse~\cite{lallart2010secearray,lefeuvre2005piezoelectric}. This architecture can achieve over 80\% extraction efficiency without continuous electrical loading or active load tuning. Its simplicity, robustness under variable excitation, and compatibility with digital controllers make it ideal for integration with adaptive MPPT and resonance tuning.

An alternative approach, Synchronized Switch Harvesting on Inductor (SSHI), offers higher extraction efficiency by inverting the piezo voltage before charge transfer~\cite{wang2021novel}, but introduces greater hardware complexity. 

This work selects SECE for its favorable trade‑off between performance and implementation overhead.

\subsection{Maximum Power Point Tracking (MPPT)}

MPPTs are various algorithms that track the theoretical maximum power point that can be extracted from an Energy Harvester. The MPPT algorithms can be roughly divided into Classic, Intelligent, and Hybrid. In addition to that, there are certain optimisation techniques such as Cuckoo Search \cite{bollipo2020hybrid}. Generally, the classic algorithms are less accurate than the other types, but are still widely used given their ease of implementation and low computational complexity.
In \cite{ottman2002adaptive}, the authors demonstrate a Perturb and Observe(P\&O) MPPT algorithm designed for PEH. P\&O is a widely used MPPT regarded for its simplicity. However, P\&O cannot track the true MPP because, by design, it fluctuates around the MPP. P\&O is a specific type of hill-climbing MPPT, MPPTs which try to reach the maxima by changing a small amount of a particular input like duty cycle.
A more advanced form of MPPT has been described in \cite{sher2015new}, which combines Fractional Short Circuit Current(FSSC) with P\&O.
The FSSC makes the initial adjustment of the MPP, followed by P\&O fine-tuning of the MPP.The FSSC is restarted only when the harvested power deviates beyond a certain threshold. This solves the slow tracking problem, which is inherent with small step P\&O. 

This paper takes inspiration from \cite{sher2015new} and introduces a Hybrid MPPT which is based on a pseudo Fractional Open Circuit Voltage (FOCV) and adaptive P\&O.

\subsection{Switched Capacitor Arrays (SCA) and tuning resonant frequency}

A switched capacitor array is a digitally controllable circuit that acts as a variable capacitor. The switched capacitor array (SCA) functions as an active tuning element, which determines the overall capacitance of the array by various combinations of inbuilt fixed-value capacitors with discrete values. Their performance hinges on accurate mapping between the ambient
frequency and the required capacitance values.

In \cite{brodersen2005mos}, the authors demonstrate one of the first works on SCA, where the authors built a recursive filter using SCA.
Since then, SCA has been used in various ways, such as rectifiers. 
For example, in \cite {salem2024recursive}, the authors designed an \\N-path SCA rectifier for PEH.

The SCA used in this paper essentially works as variable capacitor which is controlled digitally using Look Up tables (LUTs).

\subsection{Challenges and Gaps in Research}

While significant advancements have been made in PEH, particularly in damping, narrow bandwidth, and nonlinearity of PEH, gaps persist:

\begin{itemize}
    \item Lack of integrated models: Very few studies have integrated SECE, MPPT, and SCA in a single robust framework. Most studies tend to focus on dealing with problems one at a time.
    \item Limited research on using SCA to tune resonant frequency: SCA has been used before as a rectifier in PEH, but most papers do not exploit it to tune resonant frequency. 
    \item Nonlinear modeling gap: Non‑linear transducer dynamics (hysteresis) are often ignored or simplified.

\end{itemize}

\section{Methodology}

The adaptive piezoelectric energy harvesting system is modelled and validated in MATLAB using an event‑driven simulation framework based on the ode15s solver, which handles stiff and discontinuous switched‑mode power electronics effectively - the framework initializes all physical and controller parameters, then, at each iteration, alternates between SECE off‑phase integration, peak detection and charge extraction, SCA‑based resonance tuning, and hybrid MPPT updates (refer to Figure 1 for the overall simulation flow). The overall architecture comprises four interconnected subsystems: a nonlinear piezoelectric transducer, an SECE circuit, an adaptive resonance tuning network using a switched‑capacitor array, and a power management stage with a hybrid MPPT algorithm.

\begin{figure}[h]
    \centering
    \includegraphics[width=0.65\linewidth]{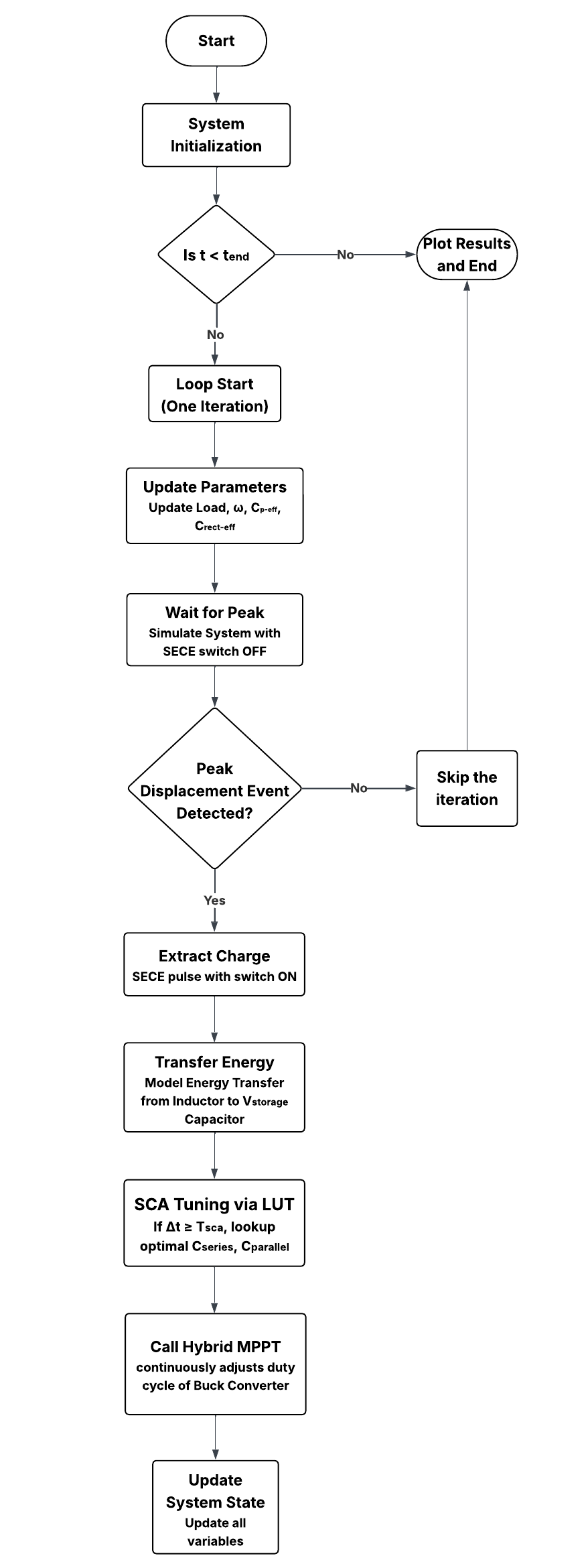}
    \caption{ A schematic representation of the entire framework.}
\end{figure}

\subsection{Nonlinear Piezoelectric Transducer Model}
The piezoelectric cantilever is represented as a second‑order electromechanical system with Bouc–Wen hysteresis to capture material memory effects under large strain.

\subsubsection{Mechanical Domain}
The governing equation of motion is:
\begin{equation}
  m\,\ddot z + c\,\dot z + F_{\mathrm{spring}}(z,h) + \alpha\,V_p = F_{\mathrm{drive}}(t)
\end{equation}
where \(m\) is the effective mass, \(c\) the damping coefficient, \(z\) the displacement, \(V_p\) the piezo voltage, and \(\alpha\) the coupling coefficient. The nonlinear spring force \(F_{\mathrm{spring}}\) depends on \(z\) and the hysteresis state \(h\) \cite{boucwen2009nonlinear}.

The nonlinear restoring force is decomposed as:
\[
  F_{\mathrm{spring}}(z,h) \;=\; \alpha\,k_s\,z \;+\; (1 - \alpha)\,k_s\,h,
\]
where \(k_s\) is the nominal stiffness and \(h(t)\) is an internal hysteretic variable that evolves according to the Bouc–Wen differential equation \cite{boucwen2009nonlinear}:
\[
  \dot h = A\,\dot z
         \;-\;\beta\,\lvert \dot z\rvert\,\lvert h\rvert^{\,n-1}\,h
         \;-\;\gamma\,\dot z\,\lvert h\rvert^n.
\]
Here, 
\(\alpha\in[0,1]\) partitions the force between purely elastic and hysteretic components, and \((A,\beta,\gamma,n)\) shape the width and smoothness of the hysteresis loop.

\subsubsection{Electrical Domain}
Piezoelectric current is described by:
\begin{equation}
  I_{\mathrm{piezo}} = \alpha\,\dot z - C_p\,\dot V_p
\end{equation}
with \(C_p\) denoting the intrinsic capacitance \cite{boucwen2009nonlinear}. During long simulation periods slight drift was observed but deemed acceptable.

\subsection{Synchronous Electric Charge Extraction}
SECE operation is divided into two event‑triggered phases at each zero‑velocity instant of the beam \cite{li2014selfsensing,lallart2010secearray}:
\begin{itemize}
  \item Phase 1: Switch open, \(I_{\mathrm{piezo}}\) charges the intermediate capacitor \(C_{\mathrm{rect}}\) while the beam vibrates freely.
  \item Phase 2: Switch closes briefly at displacement peak, forming an LC resonance between \(C_{\mathrm{rect}}\) and inductor \(L_{\mathrm{sece}}\), transferring charge to \(C_{\mathrm{storage}}\).
\end{itemize}
Some step‑size adjustments were required in early simulations \cite{li2014selfsensing}.

\subsection{Adaptive Resonance Tuning via Predictive Look-Up Table}
To overcome the narrow bandwidth, this simulation employs a predictive tuning strategy based on a look-up table (LUT). This approach models an idealized tuner to determine the maximum performance potential of the system. The tuning mechanism operates as follows:
\begin{itemize}
    \item At periodic intervals, the controller is provided with the instantaneous driving frequency, simulating a perfect frequency sensing module.
    \item This known frequency is used to query a pre-characterized look-up table, which is implemented as a linear interpolant.
    \item The LUT returns the pre-determined, optimal effective capacitance value ($C_{\mathrm{effective}}$) that maximizes power transfer for that specific frequency.
    \item This ideal capacitance value is then used directly in the system's governing electrical equations for the subsequent simulation interval.
\end{itemize}
This predictive method bypasses the challenge of real-time hill-climbing algorithms and allows for a direct evaluation of the benefits of perfect resonance matching.

\subsection{Power Management and MPPT}
Energy from \(C_{\mathrm{storage}}\) feeds a buck converter modelled with state variables for inductor current \(i_{L}\) and output voltage \(V_{\mathrm{out}}\) \cite{ottman2002adaptive}.

A hybrid MPPT controller combines:
\begin{enumerate}
  \item Adaptive P\&O with variable step size \cite{esram2007pomppt}.
  \item FOCV jumps when harvested power drops below the previous peak \cite{kim2015hybridmppt}.
\end{enumerate}

In this implementation, the true open‑circuit voltage \(V_{\mathrm{oc}}\) of the piezoelectric module is approximated by the storage capacitor voltage \(V_{\mathrm{storage}}\). This “pseudo‑FOCV” simplifies hardware requirements—eliminating the need for a dedicated open‑circuit measurement stage—and ensures the MPPT loop remains continuously active. Any small discrepancy between \(V_{\mathrm{storage}}\) and the actual \(V_{\mathrm{oc}}\) is inherently corrected by the P\&O algorithm: since P\&O continuously adjusts load conditions to maximize power, it will automatically rectify residual errors in the voltage approximation. Also, unlike a classic FOCV method that must interrupt harvesting to open‑circuit the device for voltage sampling, this approach keeps the circuit online at all times, avoiding loss of energy during measurement intervals and improving net harvested power.

\begin{figure}[h]
    \centering
    \includegraphics[width=0.65\linewidth]{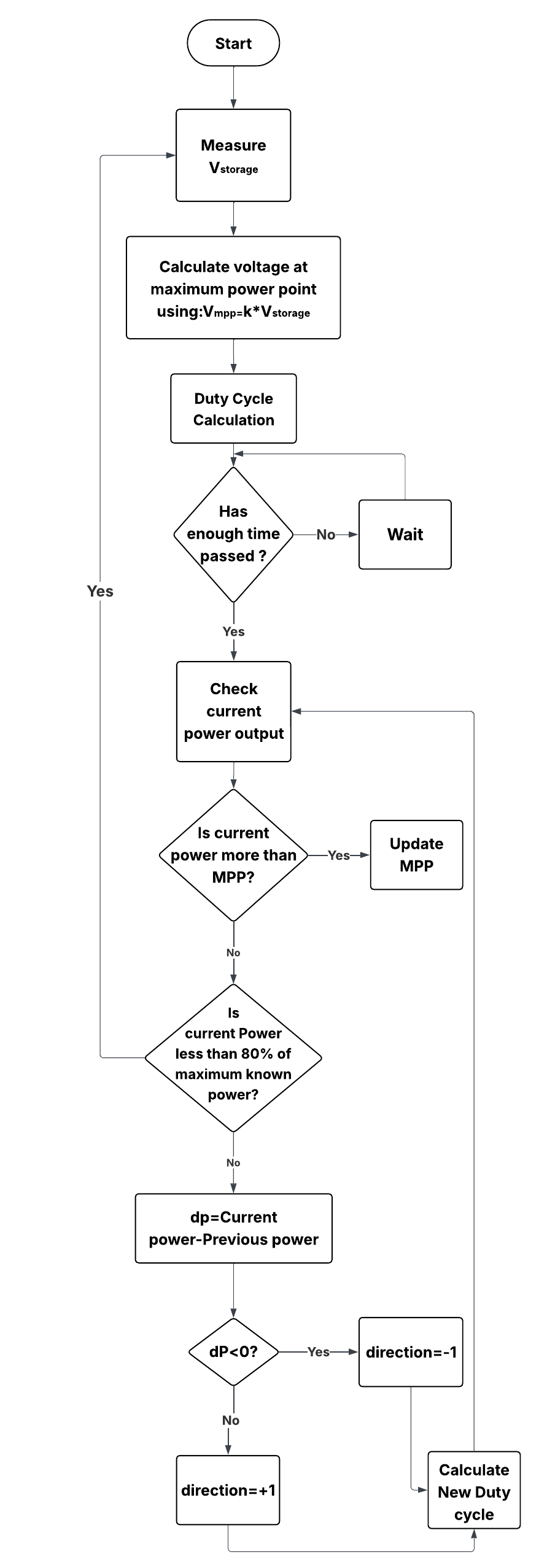}
    \caption{This figure shows the step-by-step working of the proposed MPPT.}
    \label{fig:enter-label1}
\end{figure}

Figure~\ref{fig:enter-label1} illustrates the iterative logic—measuring the storage voltage, computing a target MPP voltage, enforcing timing constraints, applying the FOCV jump when needed, and then executing the hill‑climb with directional logic and step‑size adaptation.

\section{Experimental Setup}
The numerical experiments are carried out in MATLAB to assess the proposed system's performance.

\subsection{Simulation Parameters}
Table I lists the main physical and electrical settings, selected to reflect a realistic low‐frequency piezoelectric scenario. The simulation employs ode15s solver with relative and absolute tolerances of 1e-5 and 1e-8, respectively, and no fixed maximum step size.

\begin{table}[h]
\centering
\caption{Key Simulation Parameters}
\label{tab:sim_params}
\renewcommand{\arraystretch}{1}
\begin{tabular}{l c c}
\hline\hline
\textbf{Parameter} & \textbf{Value} & \textbf{Unit} \\
\hline
Piezoelectric Transducer & & \\
\quad Mass                    & 0.01      & kg   \\
\quad Stiffness           & 4000      & N/m  \\
\quad Damping                & 0.5       & Ns/m \\
\quad Capacitance        & 15        & nF   \\
\hline
SECE Circuit & & \\
\quad Inductor         & 2.7       & mH   \\
\quad Rectifier capacitor   & 10        & nF   \\
\hline
SCA Tuning & & \\
\quad Discrete capacitor options  & [2, 4, 8, 16, 32] & nF   \\
\hline
Power Management & & \\
\quad Storage capacitor  & 100       & \textmu F \\
\quad Load resistance        & 50k      & \(\Omega\) \\
\quad Buck inductor        & 1         & mH   \\
\hline
Simulation settings & & \\
\quad Duration              & 5         & s    \\
\quad Solver tolerances (Rel/Abs) & 1e‑5/1e‑8 & ‑    \\
\hline\hline
\end{tabular}
\end{table}

\subsection{Vibration Profile}
To test the system under more realistic conditions, the harvester is subjected to a complex vibration profile. The driving frequency follows a non-linear path over the 5 second simulation, varying between 90\,Hz and 110\,Hz. Furthermore, a smoothed, band-limited random noise signal is superimposed on the primary sinusoidal driving force to simulate ambient noise and unpredictable disturbances.

\subsection{Comparative Test Cases}
Two scenarios share the same complex vibration profile:
\begin{enumerate}
    \item Control (Non‐Adaptive): The SCA is held fixed at a nominal capacitance, representing a conventional PEH tuned to a single frequency.
    \item Adaptive Test: The predictive, look-up-table-based tuning is enabled, allowing the system's effective capacitance to be updated in real-time to track the optimal resonant point as the driving frequency changes.
\end{enumerate}

\subsection{Performance Metrics}
The primary metric is total net energy delivered, found by integrating instantaneous power ($V_{\mathrm{out}}^2 / R_{\mathrm{load}}$) over 5s. Additional observations include frequency‐tracking plots to verify controller action and storage/output voltage traces to ensure power‐management stability. Overall results show improved energy capture, though minor oscillations persist in some runs.

\section{Results}
The framework was tested at a constant frequency of 100 Hz and a variable frequency from 90-110 Hz.

\begin{figure}[h]
    \centering
    \includegraphics[width=1\linewidth]{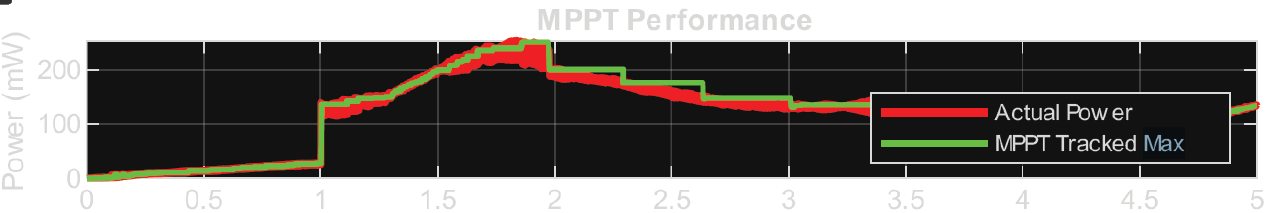}
    \caption{This plot depicts the tracking efficiency and performance of the proposed Hybrid MPPT when the frequency is time varying.}
    \label{mppt}
\end{figure}

In Figure \ref{mppt}, the performance of the Hybrid-MPPT can be observed. It can be inferred that the MPPT can successfully track the MPP when the system fluctuates between 90 Hz- 110 Hz, as shown in the plot. There is also a load change at exactly 1 second from $50\,\text{k}\Omega$ to $10\,\text{k}\Omega$
 to verify the robustness of the MPPT. The MPPT exhibits an efficiency of 93\%-95\%(subject to non-linearity).

\begin{figure}[h]
    \centering
    \includegraphics[width=1\linewidth]{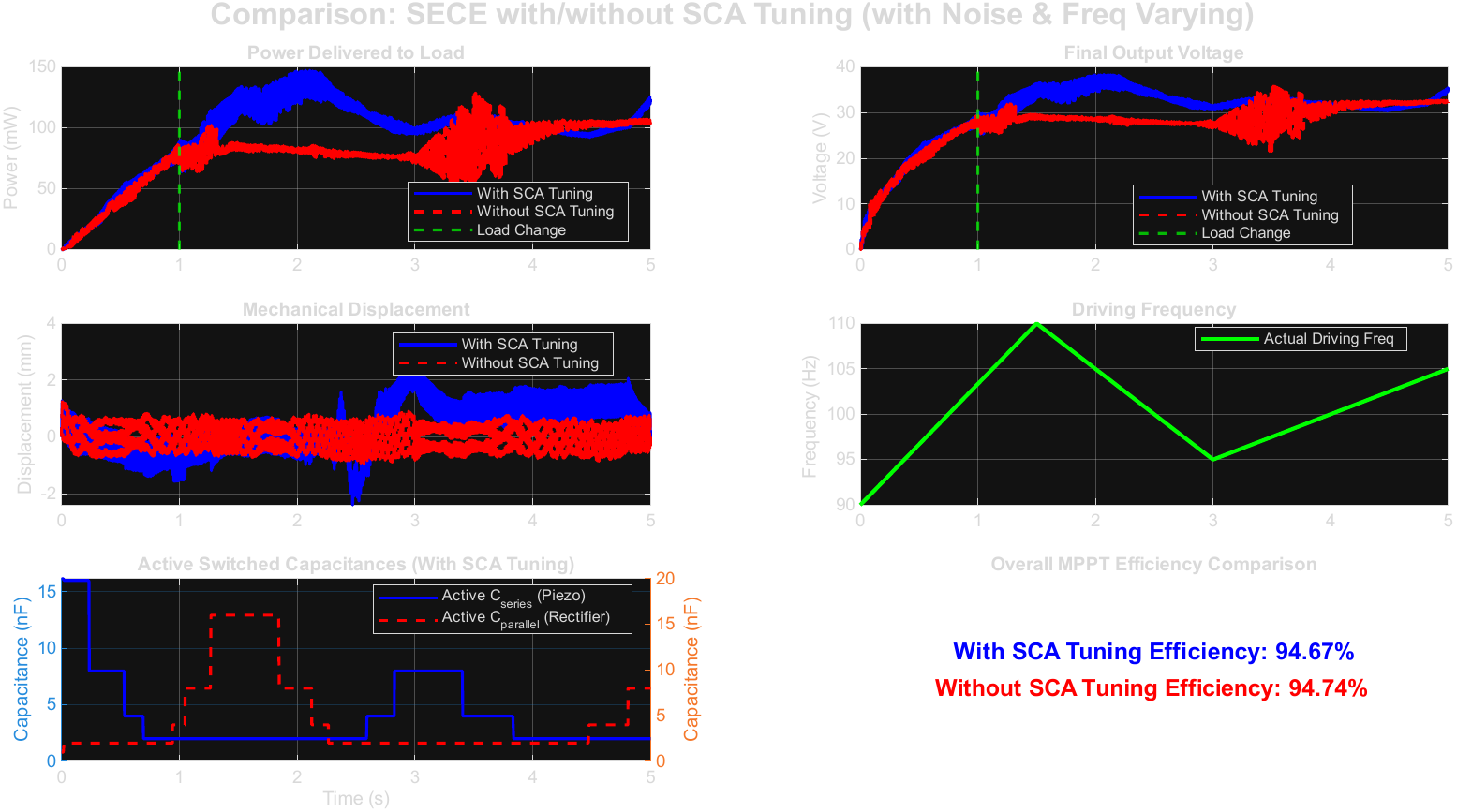}
    \caption{The above plots compare the performance of the proposed model (blue) with standard SECE (red) under variable frequency.}
    \label{comp}
\end{figure}

As seen in Figure \ref{comp}, a comparison is made between SECE only harvesting and SECE+SCA harvesting. The SECE has been implemented based on \cite{lefeuvre2005piezoelectric}. In the plots, it can be observed that when the ambient frequency deviates from 100 Hz, the SCA is able to track and tune the piezo successfully. There's a maximum of 63\% increase in harvested power when the ambient frequency is completely out of tune.  An important point to be noticed is that when the frequency falls below a defined resonance, the series SCA gets activated to match it; on the other hand, the parallel SCA gets activated when the ambient frequency becomes more than the resonance frequency. It is to be noted that using SCA, which is controlled by LUTs, is computationally more complex than simple SECE. The efficiency of the MPPT can also be observed in the bottom right corner. The slight difference in both cases can be attributed to non-linearity and computational overhead.

\begin{figure}[h]
    \centering
    \includegraphics[width=1\linewidth]{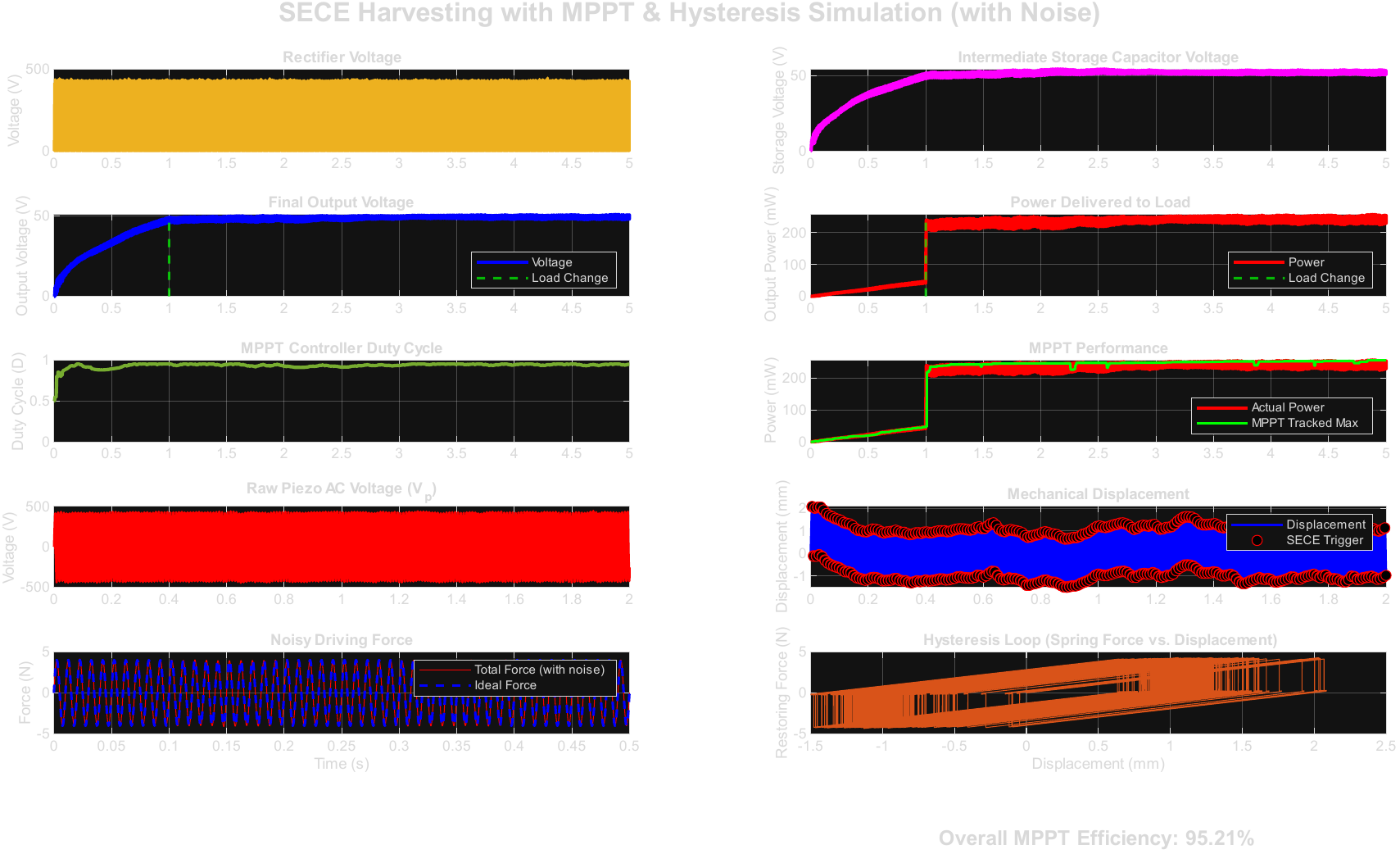}
    \caption{The above plots illustrate the performance of the entire framework at a constant frequency of 100Hz}
    \label{hystl}
\end{figure}
In the first plot of Figure \ref{hystl}, the rectifier voltage has been plotted, which averages around 350-400 volts. The next two plots show the storage capacitor voltage and final output voltage across the load, both of which average around 50 volts after the load change. The power delivered to the load is plotted next, which averages around 225 mW, followed by the MPPT controller duty cycle.
The succeeding plot exhibits the MPPT performance. an important point to be noted is that, compared to Figure \ref{mppt}, this plot exhibits the performance at a constant frequency of 100Hz. The next two plots depict raw piezo voltage and mechanical displacement, along with points where the charge is extracted and SECE is activated, and SECE triggers.

To better mimic real-world scenarios more accurately, the Bouc-Wen model of hysteresis has been implemented in this system. The actual force wave compared to the ideal sinusoidal force wave is shown in the penultimate plot. The hysteresis loop can be observed in the last plot of Figure \ref{hystl}. The area of the loop is the amount of energy lost each cycle to non-linearity.

\section{Conclusion}
The proposed model addresses key issues in PEH, including damping, narrow bandwidth, and nonlinearity. The hybrid MPPT shows significant improvement over conventional P\&O while remaining computationally simple. The Bouc–Wen model captures hysteresis losses, SECE reduces damping, and SCA enables adaptability to frequency fluctuations.

While the simulation demonstrates robust performance, real-world implementation may face challenges. The LUT-based SCA tuning assumes perfect frequency knowledge, which practical systems must estimate using phase-locked loops or predictive algorithms. Analog–digital interface delays and capacitor switching overhead could also affect efficiency.

Future work will focus on hardware implementation and LUT optimization. This study contributes toward developing adaptive and reliable PEH systems.

\section*{Acknowledgment}
We would like to thank Mars Rover Manipal, an interdisciplinary student team of MAHE, for providing the resources needed for this project.

\bibliographystyle{unsrt}
\bibliography{bibtex_reference}

\begin{thebibliography}{10}

\bibitem{pradeesh2022review}
EL~Pradeesh, S~Udhayakumar, MG~Vasundhara, and GK~Kalavathi.
\newblock A review on piezoelectric energy harvesting.
\newblock {\em Microsystem Technologies}, 28(8):1797--1830, 2022.

\bibitem{sodano2004review}
Henry~A Sodano, Daniel~J Inman, and Gyuhae Park.
\newblock A review of power harvesting from vibration using piezoelectric materials.
\newblock {\em Shock and Vibration Digest}, 36(3):197--206, 2004.

\bibitem{lefeuvre2005piezoelectric}
Elie Lefeuvre, Adrien Badel, Claude Richard, and Daniel Guyomar.
\newblock Piezoelectric energy harvesting device optimization by synchronous electric charge extraction.
\newblock {\em Journal of intelligent material systems and structures}, 16(10):865--876, 2005.

\bibitem{wang2021novel}
Xiudeng Wang, Yinshui Xia, Ge~Shi, Huakang Xia, Mengjie Chen, Zhidong Chen, Yidie Ye, and Libo Qian.
\newblock A novel mppt technique based on the envelope extraction implemented with passive components for piezoelectric energy harvesting.
\newblock {\em IEEE Transactions on Power Electronics}, 36(11):12685--12693, 2021.

\bibitem{zhang2011piezoelectric}
Shujun Zhang and Fapeng Yu.
\newblock Piezoelectric materials for high temperature sensors.
\newblock {\em Journal of the American Ceramic Society}, 94(10):3153--3170, 2011.

\bibitem{williams1996analysis}
Connel~B Williams and Rob~B Yates.
\newblock Analysis of a micro-electric generator for microsystems.
\newblock {\em sensors and actuators A: Physical}, 52(1-3):8--11, 1996.

\bibitem{feenstra2008energy}
Joel Feenstra, Jon Granstrom, and Henry Sodano.
\newblock Energy harvesting through a backpack employing a mechanically amplified piezoelectric stack.
\newblock {\em Mechanical Systems and Signal Processing}, 22(3):721--734, 2008.

\bibitem{tang2010toward}
Lihua Tang, Yaowen Yang, and Chee~Kiong Soh.
\newblock Toward broadband vibration-based energy harvesting.
\newblock {\em Journal of intelligent material systems and structures}, 21(18):1867--1897, 2010.

\bibitem{jia2019review}
Yu~Jia et~al.
\newblock Review of nonlinear vibration energy harvesting: Duffing, bi-stability, parametric, stochastic and others.
\newblock {\em Journal of Intelligent Material Systems and Structures}, 30(13):1803--1830, 2019.

\bibitem{boucwen2009nonlinear}
Guo Song, Qingsong Wang, Peng Li, and Zhiwei Li.
\newblock Nonlinear hysteresis modeling of piezoelectric actuators using a generalized bouc–wen model.
\newblock {\em Smart Materials and Structures}, 18(9):095001, 2009.

\bibitem{lallart2010secearray}
Micka{\"e}l Lallart and Daniel Guyomar.
\newblock An sece array of piezoelectric energy harvesting devices.
\newblock {\em Smart Materials and Structures}, 19(5):055015, 2010.

\bibitem{bollipo2020hybrid}
Ratnakar~Babu Bollipo, Suresh Mikkili, and Praveen~Kumar Bonthagorla.
\newblock Hybrid, optimal, intelligent and classical pv mppt techniques: A review.
\newblock {\em CSEE Journal of Power and Energy Systems}, 7(1):9--33, 2020.

\bibitem{ottman2002adaptive}
Geffrey~K Ottman, Heath~F Hofmann, Archin~C Bhatt, and George~A Lesieutre.
\newblock Adaptive piezoelectric energy harvesting circuit for wireless remote power supply.
\newblock {\em IEEE Transactions on power electronics}, 17(5):669--676, 2002.

\bibitem{sher2015new}
Hadeed~Ahmed Sher, Ali~Faisal Murtaza, Abdullah Noman, Khaled~E Addoweesh, Kamal Al-Haddad, and Marcello Chiaberge.
\newblock A new sensorless hybrid mppt algorithm based on fractional short-circuit current measurement and p\&o mppt.
\newblock {\em IEEE Transactions on sustainable energy}, 6(4):1426--1434, 2015.

\bibitem{brodersen2005mos}
Robert~W Brodersen, Paul~R Gray, and David~A Hodges.
\newblock Mos switched-capacitor filters.
\newblock {\em Proceedings of the IEEE}, 67(1):61--75, 2005.

\bibitem{salem2024recursive}
Loai~G Salem.
\newblock A recursive $ n $-path switched-capacitor rectifier for piezoelectric energy harvesting.
\newblock {\em IEEE Journal of Solid-State Circuits}, 2024.

\bibitem{li2014selfsensing}
Hongliang Li, Micka{\"e}l Lallart, and Daniel Guyomar.
\newblock A self-sensing synchronous electric charge extraction (sece) solution for piezoelectric energy harvesting enhancement.
\newblock {\em IEEE Transactions on Power Electronics}, 29(3):1244--1252, 2014.

\bibitem{esram2007pomppt}
Trishan Esram and Patrick~L Chapman.
\newblock Advanced perturb and observe algorithm for maximum power point tracking in photovoltaic systems with adaptive step size.
\newblock {\em IEEE Transactions on Energy Conversion}, 22(2):439--449, 2007.

\bibitem{kim2015hybridmppt}
Jaehoon Kim and Jang-Myung Kim.
\newblock Design and analysis of a hybrid mppt method for pv systems.
\newblock {\em IEEE Transactions on Power Electronics}, 30(4):2042--2054, 2015.

\end{thebibliography}

\end{document}